\documentclass[final,5p,times]{elsarticle}
\usepackage{amssymb}
\journal{}

\begin{document}

\begin{frontmatter}

\title{High Temperature Superconductivity from Strong Correlation
}

\author{Takashi Yanagisawa} 

\address{Electronics and Photonics Research Institute, 
National Institute of Advanced Industrial Science and Technology (AIST),
Central 2, 1-1-1 Umezono, Tsukuba, Ibaraki 305-8568, Japan
}
%\email{t-yanagisawa@aist.go.jp}

\begin{abstract}
It is important to understand the mechanism of high-temperature 
superconductivity.  It is obvious that the interaction with large energy 
scale is responsible for high critical temperature $T_c$.  
The Coulomb interaction is one of candidates that bring about high-temperature 
superconductivity because its characteristic energy is of the order of eV.  
There have been many works for the Hubbard model including three-band 
d-p model with the on-site Coulomb repulsion to investigate a possibility 
of high-temperature superconductivity.  It is, of course, not trivial 
whether the on-site Coulomb interaction leads to a pairing interaction 
between two electrons.  We argue that high-temperature superconductivity 
is possible in the strongly correlated region by using the variational 
Monte Carlo method for the two-dimensional t-U-J-V model.  The exchange 
interaction J and the nearest-neighbour attractive interaction V cooperate 
with U and will act to enhance the critical temperature.
\end{abstract}


\end{frontmatter}

\section{Introduction}

A basic question in the study of superconductivity is on an upper 
limit of the critical temperature $T_c$.  
It has been suggested that the maximum of $T_c$ is in the range from 
30K to 40K when the superconducting transition is based on the 
electron-phonon interaction [1, 2].  
The prime reason of low $T_c$ for the electron-phonon mechanism is 
that the Debye frequency $\omega_D$ is of the order of 100K and 
$T_c$ is reduced 
by more than one digit than $\omega_D$.  In general, the electron-phonon 
coupling constant $\lambda$ is small.  
Thus there is an upper limit for $T_c$ within the electron-phonon mechanism.  
The interaction of large energy scale will be responsible for high 
temperature superconductivity.  The electronic origin mechanisms of 
superconductivity are worth studying in the search of high temperature 
superconductors.  The electronic interaction comes from the Coulomb 
interaction between electrons.  
From this point of view, it is important to consider the intra-atomic 
Coulomb interactions that are of the order of eV.  Then a question 
arises as to whether the superconductivity is indeed induced by the 
Coulomb interaction.  This subject has been studied intensively by 
using electronic models with the on-site Coulomb interaction $U$ 
including the two-dimensional (2D) Hubbard model [3-26], the d-p 
model [27-33] and also the ladder Hubbard model [34-38].

It remains matter of controversial as to whether the 2D Hubbard model 
accounts for high-temperature superconductivity.  Most of the results 
of quantum Monte Carlo methods have failed to show the existence of 
superconductivity [13,14,18], and some results, however, still keep 
open the possibility of superconductivity [15, 24].  
We must note that quantum Monte Carlo methods are methods being 
applicable only in the weakly correlated region where $U/t$ is not 
large and is at most from 4 to 5.  The existence of superconductivity 
in strongly correlated regions should be considered further.  
This may account for properties of high-temperature cuprate superconductors.

\section{Superconductivity in Strongly Correlated Region}

Superconductivity in the strongly correlated region is described by 
using the 2D Hubbard model.  The Hamiltonian is given by
\begin{equation}
H= \sum_{ij}t_{ij}c^{\dag}_{i\sigma}c_{j\sigma}
+U\sum_i n_{i\uparrow}n_{i\downarrow}.
\end{equation}
$\{t_{ij}\}$ are transfer integrals and $U$ is the on-site Coulomb energy.
The transfer integral $t_{ij}$ is non-zero $t_{ij} = −t$ for nearest-neighbor 
pair $\langle ij \rangle$ and $t_{ij} = −t’$ for next-nearest neighbor pair
$\langle ij\rangle$; otherwise $t_{ij}$ vanishes.  
We denote the number of sites as $N$ and the number of electrons as $N_e$.

The ground state of the Hubbard model is divided into two regions, 
that is, a weak correlation region and a strong correlation region.  
When the Coulomb interaction $U$ is as large as or greater than the 
bandwidth $8t$, the ground state should be regarded as the strongly 
correlated one [39, 40].  We use the Gutzwiller ansatz and evaluate the 
superconducting condensation energy Econd for d-wave pairing.  
The wave function is
\begin{equation}
\psi = P_{N_e}P_G\psi_{BCS},
\end{equation}
where $P_G$ is the Gutzwiller operator
$P_G=\prod_j(1-(1-g)n_{j\uparrow}n_{j\downarrow})$ with the parameter
$g$ in the range of $0\le g\le 1$.
The gap function $\Delta$ in the BCS function $\psi_{BCS}$ is regarded
as a variational parameter.
$P_{N_e}$ is a projection operator that extracts only the state with a 
fixed total electron number $N_e$.  The condensation energy is defined as
$E_{cond}= E(\Delta=0)-E(\Delta=\Delta_{opt})$
for the optimized gap function $\Delta_{opt}$.

We show Econd as a function of $U$ in Fig.1.  $E_{cond}$ increases rapidly 
near $U\sim 8t$ as going into the strongly correlated region.  
$E_{cond}$ is very small in the weakly correlated region for $U < 8t$ 
and almost vanishes for $t’ = 0$.  For such small value of $E_{cond}$, 
it is certainly hard to obtain a signal of superconductivity by means of 
numerical methods such as the quantum Monte Carlo method.  
It follows from Fig.1 that high temperature superconductivity is possible 
only in the strongly correlated region.  In particular, $U/t\sim 10-14$ 
is favorable for superconductivity.  Similar results have been reported for 
$t’/t= −0.1$, $−0.25$ and $−0.4$ in Ref. [26].

\begin{figure}
\begin{center}
\includegraphics[width=9cm]{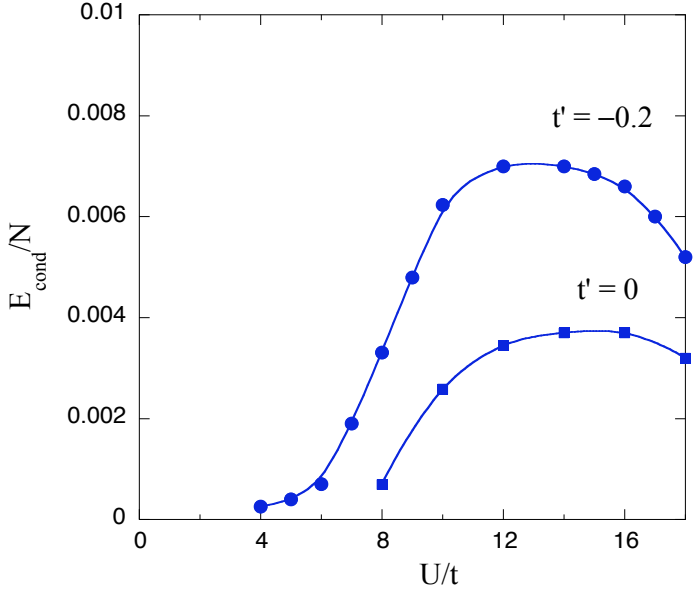}
\caption{
The superconducting condensation energy per site as a function of $U$
in units of $t$ for $t’ = 0$ and $t’ = −0.2$ on $10\times 10$ lattice.  
The number of electrons is $N_e = 88$ for $t’ = 0$ and $N_e = 84$ for 
$t’ = −0.2$.
}
\end{center}
\label{fig1}
\end{figure}

\section{Enhancement of Superconductivity with Weak Attractive
Interactions}

Let us examine when the critical temperature $T_c$ is enhanced in 
correlated-electron systems.  A layered crystal structure is obviously 
an important factor.  The increase of the density of states in a 
layered crystal is certainly important and an interlayer interaction 
plays a significant role in finding new superconductors [41-45].  
Apart from this, if there is an interaction that promotes superconductivity 
cooperating with the Coulomb interaction, $T_c$ will be increased.  
The $d$-wave pairing state in the Hubbard model will be more stabilized 
by weak nearest-neighbor attractive interactions.  
We consider the exchange interaction $J$ and attractive interaction $V$ 
given as
\begin{eqnarray}
H&=& \sum_{ij}t_{ij}c^{\dag}_{i\sigma}c_{j\sigma}
+U\sum_i n_{i\uparrow}n_{i\downarrow}
+J\sum_{\langle ij\rangle}\left( {\bf S}_i\cdot{\bf S}_j
-\frac{1}{4}n_in_j \right)\nonumber\\
&& +V\sum_{\langle ij\rangle}n_in_j.
\end{eqnarray}
This is so called the t-U-J-V model.  
In high-temperature cuprates, the exchange coupling $J$ arises as a 
super-exchange interaction between d electrons in copper atoms, which 
is mediated by oxygen atoms.  The last term with negative $V$ is expected 
to arise from the electron-phonon interaction [46].  
In the variational Monte Carlo calculations, both $J$ and $V$ ($< 0$) act 
as an attractive interaction for nearest-neighbour pairs of electrons.  
The Gutzwiller ansatz is also adopted here.  We use the electron-hole 
transformation to fix the electron number in $\psi_{BCS}$ instead of 
using $P_{N_e}$ [47].  We show the results in Figs.2 and 3.  
In Fig.2 we compare $−E_{cond}$ for $J/t$ = 0, 0.1 and 0.2 where 
$U/t = 12$ and $t’/t = −0.2$.  Basically the exchange coupling $J$ works 
effectively in a similar way to $U$ to form the d-wave pairing.  
We obtain a similar result for the interaction $V$, where the 
lowering of the energy $−E_{cond}$ by $V$ is smaller than that by $J$.  
It should be noted, however, that $J$ alone can not stabilize the 
d-wave pairing, which means that if $U$ is small, the d-wave 
superconducting state is not realized even for $J > 0$.  This is shown 
in Fig.3 where $−E_{cond}$ is shown as a function $\Delta$ with 
$J = 0.1$ and $t’ = 0$ for $U/t = 8$ and 12.  $E_{cond}$ is very small 
for $U/t = 8$ and vanishes for small $U$.  This indicates that the 
strong on-site correlation is needed so that the exchange coupling $J$ 
induces superconductivity.

\begin{figure}
\begin{center}
\includegraphics[width=9cm]{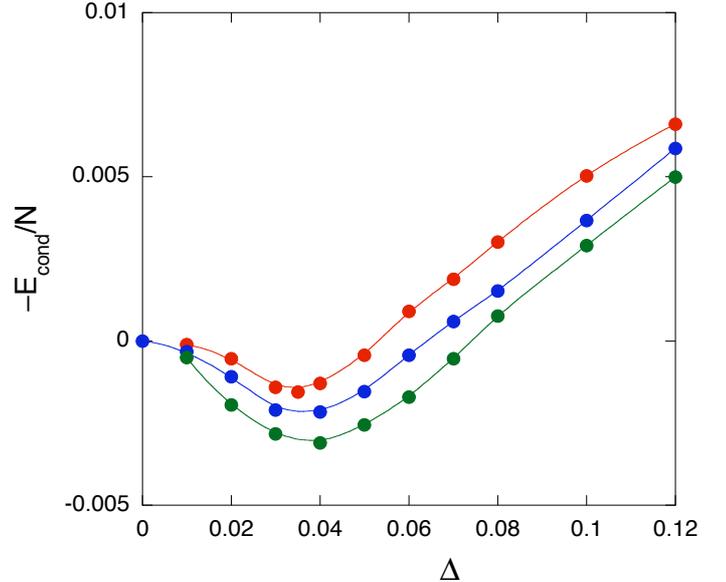}
\caption{
The superconducting condensation energy per site as a function 
of $\Delta$ in units of $t$.
From the top, we set $J$ = 0, 0.1 and 0.3 for parameters $U$ = 12 and 
$t’$ = −0.2.  The number of electron is i$N_e$ = 84 on 
$10\times 10$ lattice.
}
\end{center}
\end{figure}

\begin{figure}
\begin{center}
\includegraphics[width=9cm]{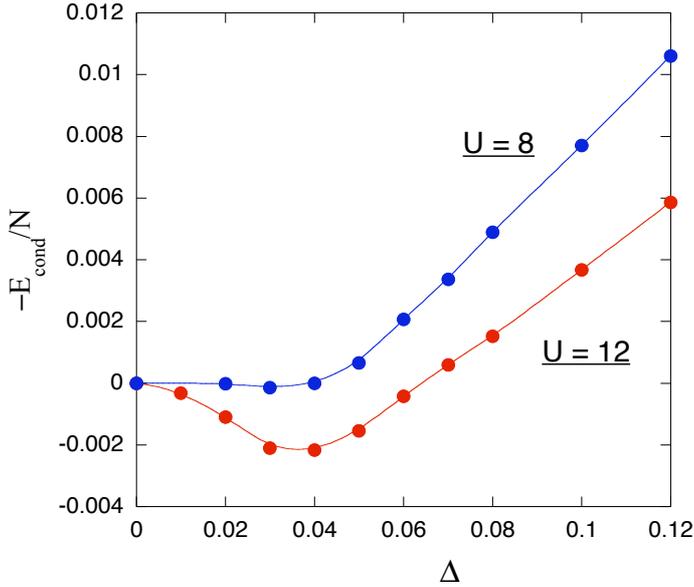}
\caption{
The superconducting condensation energy per site as a function 
of $\Delta$ in units of $t$.
The condensation energy as a function of $\Delta$ in units of $t$ for 
$J$ = 0.1 and $t’$ = 0 where $U$ = 8 and $U$ = 12.
}
\end{center}
\end{figure}

\section{Summary}

We have investigated the two-dimensional t-U-J-V Hubbard model.  
The condensation energy becomes large in the strongly correlated region, 
suggesting a possibility of high-temperature superconductivity.  
We mention here that there is a report that suggests a high-temperature 
superconductivity in the moderate-$U$ region by means of the dynamical 
mean field theory [25].  We must further clarify a consistency with the 
results obtained by the quantum Monte Carlo method.

The exchange interaction $J$ and the nearest-neighbor attractive 
interaction $V$ cooperate with the Coulomb interaction $U$ to enhance 
superconductivity.  It is important that the exchange coupling $J$ alone 
cannot promote superconductivity if $U$ is small in the weakly 
correlated region.  We expect that the strong $U$ and weak $J$ and $V$ 
may realize high critical temperature

\section*{Acknowledgments}
The author expresses his sincere thanks to K. Yamaji, M. Miyazaki, 
I. Hase and S. Koikegami for valuable discussions.  
A part of computations was performed at the facility of supercomputer 
center of the Institute for Solid State Physics, the University of Tokyo.
This work was supported by Grant-in-Aid for Scientific Research from the
Ministry of Education, Culture, Sports, Science and Technology in Japan.


\vspace{1cm}

\begin{thebibliography}{9}

\bibitem{mcm68}
W. L. McMillan, Phys. Rev. 167 (1968) 331.
\bibitem{all75}
P. B. Allen, R. C. Dynes, Phys. Rev. B12 (1975) 905.
\bibitem{hir85}
J. E. Hirsch, Phys. Rev. B31 (1985) 4403.
\bibitem{yok88}
H. Yokoyama, H. Shiba, J. Phys. Soc. Jpn. 57 (1988) 2482.
\bibitem{whi89}
S. R. White, D. J. Scalapino, R. L. Sugar, E. Y. Loh, J. E. Gubernatis, 
R. T. Scalettar, Phys. Rev. B40 (1989) 506.
\bibitem{hir89}
J. E. Hirsch, D. Loh, D. J. Scalapino, S. Tang. Phys. Rev. B39 (1989) 243.
\bibitem{sca89}
R. T. Scalettar, D. J. Scalapino, R. L. Sugar, S. R. White. Phys. Rev. B39 (1989) 243.
\bibitem{mor92}
A. Moreo, Phys. Rev. B45 (1992) 5059.
\bibitem{yan96}
T. Yanagisawa, Y. Shimoi, Int. J. Mod. Phys. B10 (1996) 3383.
\bibitem{nak97}
T. Nakanishi, K. Yamaji and T. Yanagisawa, J. Phys. Soc. Jpn. 66 (1997) 294.
\bibitem{yam98}
K. Yamaji, T. Yanagisawa, T. Nakanishi and S. Koike,, Physica C304 (1998) 225. 
\bibitem{yam00}
K. Yamaji, T. Yanagisawa and S. Koike, Physica B284 (2000) 415.
\bibitem{zha97}
S. Zhang, J. Carlson and J. E. Gubernatis, Phys. Rev. B55 (1997) 7464.
\bibitem{zha97b}
S. Zhang, J. Carlson and J. E. Gubernatis, Phys. Rev. Lett. 78 (1997) 4486.
\bibitem{kur97}
K. Kuroki, H. Aoki, Phys. Rev. B56 (1997) R14287.
\bibitem{bul02}
N. Bulut, Advances in Phys. 51 (2002) 1587.
\bibitem{miy04}
M. Miyazaki, T. Yanagisawa, K. Yamaji. J. Phys. Soc. Jpn. 73 (2004) 1643.
\bibitem{aim07}
T. Aimi, M. Imada, J. Phys. Soc. Jpn. 76 (2007) 113708.
\bibitem{miy09}
M. Miyazaki, K. Yamaji, Y. Yanagisawa, R. Kadono. J. Phys. Soc. Jpn. 78 (2009) 043706.
\bibitem{yam11}
K. Yamaji, T. Yanagisawa, M. Miyazaki, R. Kadono. J. Phys. Soc. Jpn. 80 (2011) 083702.
\bibitem{kon01}
J. Kondo, J. Phys. Soc. Jpn. 70 (2001) 808.
\bibitem{hlu99}
R. Hlubina, Phys. Rev. B59 (1999) 9600.
\bibitem{yan08}
T. Yanagisawa, New J. Phys. 10 (2008) 023014.
\bibitem{yan13}
T. Yanagisawa, New J. Phys. 15 (2013) 033012.
\bibitem{gul13}
E. Gull, O. Parcdlet, A. J. Millis, Phys. Rev. Lett. 110 (2013) 216405.
\bibitem{yok13}
H. Yokoyama, M. Ogata, Y. Tanaka, K. Kobayashi, H. tsuchiura, 
J. Phys. Soc. Jpn. 82 (2013) 014707.
\bibitem{eme97}
V. J. Emery, Phys. Rev. Lett. 58 (1997) 2794.
\bibitem{web09}
C. Weber, A. Lauchi, F. Mila, T. Giamarchi. Phys. Rev. Lett. 102 (2009) 017005.
\bibitem{lau11}
B. Lau, M. Berciu, G. A. Sawatzky. Phys. Rev. Lett. 106 (2011) 036401.
\bibitem{yan01}
T. Yanagisawa, S. Koike, K. Yamaji. Phys. Rev. B64 (2001) 184509.
\bibitem{yan03}
T. Yanagisawa, S. Koike, K. Yamaji, Phys. Rev. B67 (2003) 132408.
\bibitem{yan09}
T. Yanagisawa, M. Miyazaki, K. Yamaji. J. Phys. Soc. Jpn. 78 (2009) 013706.
\bibitem{yan13b}
T. Yanagisawa, M. Miyazaki, K. Yamaji. J. Mod. Phys. 4 (2013) 33.
\bibitem{noa97}
R. M. Noack, N. Bulut, D. J. Scalapino, M. G. Zacher, Phys. Rev. B56 (1997) 7162.
\bibitem{kur96}
K. Kuroki, T. Kimura, H. Aoki, Phys. Rev. B54 (1996) R15641.
\bibitem{koi99}
S. Koike, K. Yamaji, T. Yanagisawa, J. Phys. Soc. Jpn. 68 (1999) 1657.
\bibitem{yam94}
K. Yamaji K, Y. Shimoi, T. Yanagisawa, Physica C235-240 (1994) 2221.
\bibitem{yan95}
T. Yanagisawa, Y. Shimoi, K. Yamaji, Phys. Rev. B52 (1995) R3860.
\bibitem{yok06}
H. Yokoyama, M. Ogata, Y. Tanaka, J. Phys. Soc. Jpn. 95 (2006) 114906.
\bibitem{yan14}
T. Yanagisawa, M. Miyazaki, Europhys. Lett. 107 (2014) 27004.
\bibitem{mae94}
Y. Maeno, H. Hashimoto, K. Yoshida, S. Nishizaki, T. Fujita, J. G. Bednorz, 
F. Lichtenberg, Nature 372 (1994) 532.
\bibitem{heb91}
A. F. Hebard, M. J. Rosseinsky, R.C. Haddon, D. W. Murphy, S. H. Glarum, 
T. T. Palstra, A. P. Ramirez, A.R. Kortan, Nature 350 (1991) 600.
\bibitem{nag01}
J. Nagamatsu, N. Nakagawa, T. Muranaka, Y. Zenitani, J. Akimitsu, 
Nature 410 (2001) 63.
\bibitem{koi06}
S. Koikegami and T. Yanagisawa, J. Phys. Soc. Jpn. 75 (2006) 034715;
S. Koikegami and T. Yanagisawa, J. Phys. Soc. Jpn. 70 (2001) 3499; 
ibid. 71 (2002) 671 (E).
\bibitem{koi03}
S. Koikegami and T. Yanagisawa, Phys. Rev. B67 (2003) 134517.
\bibitem{har09}
T. M. Hardy, J. P. Hague, J. H. Samson, A. S. Alexandrov, 
Phys. Rev. B79 (2009) 212501.
\bibitem{yan98}
T. Yanagisawa, S. Koike, K. Yamaji, J. Phys. Soc. Jpn. 67 (1998) 3867; 
T. Yanagisawa, S. Koike, K. Yamaji, J. Phys. Soc. Jpn. 68 (1999) 3608


 
\end{thebibliography}
\end{document}